\documentclass[conference]{IEEEtran}

\usepackage[utf8]{inputenc} 
\usepackage[T1]{fontenc}
\usepackage{url}
\usepackage{ifthen}
\usepackage{cite}
\usepackage[cmex10]{amsmath} 
\usepackage{amssymb}
\usepackage{graphicx}
\usepackage{multirow}
\usepackage{array}
\usepackage{makecell}
\usepackage{tikz}
\usepackage{algorithm}
\usepackage{algpseudocode}
\usepackage{comment}
\usepackage{mathtools}
\usepackage[english]{babel}

\newtheorem{thm}{Theorem}
\newtheorem{lem}{Lemma}

\newtheorem{cor}{Corollary}

\newtheorem{defn}{Definition}

\newtheorem{exmp}{Example}

\newtheorem{rem}{Remark}

\begin{document}
	\title{Multi-access Coded Caching from a New Class of Cross Resolvable Designs} 
	
	\author{%
		\IEEEauthorblockN{Pooja Nayak Muralidhar and B. Sundar Rajan \IEEEauthorrefmark{1}}
		\IEEEauthorblockA{\IEEEauthorrefmark{1}Department of Electrical Communication Engineering, Indian Institute of Science, Bengaluru 560012, KA, India \\
			E-mail: \{poojam,bsrajan\}@iisc.ac.in}
	}
	\maketitle
	\thispagestyle{plain}
	\pagestyle{plain}
	\begin{abstract}
    Multi-access coded caching schemes from cross resolvable designs (CRD) have been reported recently \cite{KNRarXiv}. To be able to compare coded caching schemes with different number of users and possibly with different number of caches a new metric called  rate-per-user  was introduced and it was shown that under this new metric the schemes from CRDs perform better than the Maddah-Ali-Niesen scheme in the large memory regime.   In this paper a new class of CRDs is presented and it is shown that the multi-access coded caching schemes derived from these CRDs perform better than the Maddah-Ali-Niesen scheme in the entire memory regime.  Comparison with other known multi-access coding schemes is also presented. 
	\end{abstract}
	\section{INTRODUCTION}
    Coded caching is an active area of research that has gained popularity due to its ability to reduce data transmissions during the times of high network congestion by prefetching parts of demanded contents into the memories of end users.
    Designing schemes that are well suited to meet practical constraints like subpacketization while achieving reasonable rates is the main challenge in developing good coded caching schemes.
    Most of the attention has been centered around the scenario where users are equipped with dedicated caches.
      
    However, in a variety of settings, such as different cellular networks, multiple users can share a single cache or users can conceivably connect to multiple caches whose coverage areas may overlap. In such cases caching schemes suited for such situations have to be implemented. In the recent years, coded caching schemes for various cache share situations have been studied in the work of \cite{SBP2}, \cite{AAA}, \cite{HKD}, \cite{SPE}, \cite{RaK3}, \cite{KNRarXiv}, \cite{CLWZC}, \cite{BL} etc. In this work, we propose a new construction of a combinatorial structure called cross resolvable designs which was first discussed in \cite{KNRarXiv} and develop multi-access coded caching from it.
	\subsection{Multi-access Coded Caching - System Model}
	\label{sec1A}
	Fig. \ref{fig1} shows a multi-access coded caching system with a unique server $\mathcal{S}$ storing $N$ files $W_{1}$,$W_{2}$,$W_{3}$,\dots,$W_{N}$ each of unit size. There are $K$ users in the network connected via an error-free shared link to the server $\mathcal{S}.$ The set of users is denoted by $\mathcal{K}.$  There are $b$ number of helper caches each of size $M$ files. Each user has access to $z$ out of the $b$ helper caches. Let $\mathcal{Z}_k$ denote the content in the $k$-th cache. It is assumed that each user has an unlimited capacity link to the caches it is connected to. 
	
	There are two phases: the placement phase and the delivery phase. During the placement phase, certain parts of each file are stored in each cache, which is carried out during the off-peak hours. During the peak hours, each user demands a file, and the server broadcasts coded transmissions such that each user can recover its demanded file by combining the received transmissions with what has been stored in the caches it has access to. This is the delivery phase. The coded caching problem is to jointly design the placements and the delivery with a minimal number of transmissions to satisfy all users' demands. The amount of transmissions used in the unit of files is called the {\it rate} or the {\it delivery time}. Subpacketization level is the number of packets that a file is divided into. Coding gain is defined as the number of users who are benefited in a single transmission. 
	\subsection{Known Coded caching schemes with dedicated caches}
	\label{sec1B}
\begin{figure}
		\begin{center}
			\includegraphics[width=7cm,height=6cm]{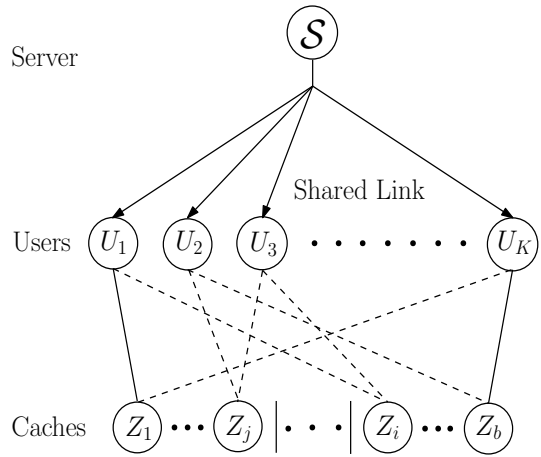}
			\caption {Problem setup for multi-access coded caching with $K$ users, $b$ caches and each user, connected to $z$ caches.}
			\label{fig1}
		\end{center}
	\end{figure}

	The framework of the seminal paper \cite{MaN} considers a network with $K$ users, each equipped with memory of size ${M}$ and $N$ files of very large size among which each user is likely to demand any one file. The rate $R$ achieved is 
	$ K  \left(1 - \frac{{M}}{N}\right) \frac{1}{(1 + K\frac{{M}}{N})}.$ The factor $(1 + K\frac{{M}}{N})$ which is originally call global caching gain is also known as the {\it coding gain} or the {\it Degrees of Freedom (DoF)}. We refer to this scheme as the MaN scheme henceforth. This original setup can be viewed as a special case of the scheme corresponding to Fig. \ref{fig1} with $b=K$ and $z=1$ which may be viewed as each user having a dedicated cache of its own. 	
	\\
	To mitigate the exponential subpacketization problem that arises, by employing the MaN scheme several other schemes based on resolvable block designs from linear block codes \cite{TaR}, hypergraphs \cite{SZG}, block designs \cite{KrP1}, schemes in \cite{CJYT} etc., have also been proposed in the recent years. These approaches have helped to reduce the subpacketization problem, by mainly trading off with respect to rate or memory fraction accessible to the user as compared to the scheme in \cite{MaN}.
	\subsection{Known Multi-access Coded Caching Schemes}
    \label{sec1C}
	The multi-access setup refers to the case when a user has access to more than one cache, and multiple users can access the same cache. This setup was first considered in the work of \cite{HKD}, 
	where a $K$ users and $K$ caches with each user having access to neighboring $z$ caches in a cyclic wraparound fashion, was dealt with. In the original paper \cite{HKD}, the scheme proposed is decentralized. However the scheme can be extended to get a centralized scheme which gives the rate given by
	$$R = \frac{K - \frac{KzM}{N}}{1 + \frac{KM}{N}} \text{ for }  M \leq \frac{N}{z}$$
	and $R = 0$ for $M > \frac{N}{z}$.
	We will refer to this scheme as the HKD scheme.
	 
	The scheme of \cite{RaK3}, also considers a multi-access setup with cyclic wraparound and for any access degree $z$ achieves the rate given by the expression\\
	\begin{align*}
	R =  
	\begin{cases}
	K \Big{(1-z\;\frac{M}{N}\Big)}^2 \text{ for } i \in {0 \cup \Big[\Bigl\lfloor \frac{K}{z} \Bigr\rfloor\Big]} 
	\text{ and }\\
	 0 \text{ for } i = \Bigl \lceil \frac{K}{z} \Bigr \rceil
	\end{cases}
	\end{align*}
	for $M = i\frac{N}{K}$ where $i \in {0 \cup [\lceil \frac{K}{z} \rceil]}$ 
	We refer to this scheme as the RK scheme.
	
	 The work of \cite{SPE} also deals with the same problem setup as in the previous cases and provides two new schemes which can serve, on average, more than $K\gamma + 1$ users at a time and for the special case of $z = \frac{K-1}{K\gamma}$, the achieved gain is proved to be optimal under uncoded cache placement where $\gamma = \frac{M}{N},\;\gamma \in \{1/K, 2/K,\dots,1\}$. 
	 The general scheme is proposed for the case $K\gamma = 2$.
	 We will refer to this scheme as the SPE scheme.
	 
	 The coded caching scheme proposed by \cite{HKD} suffers
	from the case that $z$ does not divide $K$, where the needed number of transmissions is at most four times the load
	expression for the case where $z$ divides $K$. 
	The work of \cite{CLWZC} proposes a new scheme for the same setup considered in \cite{HKD} so that the
	case where $L$ divides $K$, remains achievable in full generality and gives the same rate expression as in \cite{HKD}. 
	where $z$ is the access degree, $M$ the memory associated with each cache and $N$ the number of files. The subpacketization $F = K \binom{K'}{t}$ where $t = K \frac{M}{N}$ and $K' = K - t(z -1)$.
	The effective file fraction accessible to the user in all the topologies discussed before is $\frac{zM}{N}$.
    We will refer to this scheme as the CLWZC scheme.
    	
	In \cite{KNRarXiv}, the authors develop a multi-access scheme from a specific type of resolvable designs called Cross Resolvable Designs (CRD) and the setup of the multi-access network discussed here depends on the cross resolvable design chosen. The effective file fraction accessible to the user by this setup is less than $\frac{zM}{N}$ as opposed to the other schemes discussed previously.
	The number of users supported here, is significantly higher than the other existing schemes for the same number of caches and for practically realizable subpacketization levels. We will refer to this scheme as the KMR scheme.
	Our work considers only the setup in \cite{KNRarXiv}. 
	
	To perform comparison of coded caching schemes for various multi-access topologies, we use the metric per user rate or rate per user defined in \cite{KNRarXiv}.
	To allow a fair comparison between centralized coded caching schemes with dedicated cache and centralized multi-access schemes, we introduce the notion of access gain. It is the gain of the multi-access system over the optimal centralized coded caching scheme  with dedicated cache with uncoded placement for the same number of caches, memory fraction stored in the caches and is defined as
	$$g_{access} = \frac{\splitfrac{\text{Per user rate of optimal centralized coded caching}}{\text{scheme  with dedicated cache}}}{\text{Per user rate of multi-access system}}$$
	
\subsection{Contributions}
\label{sec1E}
	The contributions of this paper may be summarized as follows:

\begin{itemize}
\item We present a construction for a new family of cross resolvable designs (CRDs) which enables construction of multi-access coded caching schemes with parameters not achievable with the CRDs presented in \cite{KNRarXiv}.

\item When used for the construction given in \cite{KNRarXiv} the new CRDs lead to multi-access coded caching schemes with better performance. In particular the multi-access coded caching schemes presented in \cite{KNRarXiv} give smaller per user rate than the MaN scheme only in the high cache storage regime whereas the CRDs presented in this paper give smaller per user rate for all cache storage regime. 

%
		
\item The schemes from the new CRDs presented  when compared to other multi-access schemes perform better in terms of number of users supported, per user rate and subpacketization.
	\end{itemize}

	\section{BACKGROUND}
	\label{sec2}
	In this section we review some of the preliminaries from \cite{KNRarXiv}.
	\begin{defn}\cite{TaR}
		A design is a pair  $(X, \mathcal{A})$ such that
		\begin{itemize}
			\item $X$ is a finite set of elements called points, and
			\item $\mathcal{A}$ is a collection of nonempty subsets of $X$ called blocks, where each block contains the same number of points.
		\end{itemize}
	\end{defn}
	\begin{defn}\cite{TaR}
		A parallel class $\mathcal{P}$ in a design $(X, \mathcal{A})$ is a subset of disjoint blocks from $\mathcal{A}$ whose union is $X$. A partition of $\mathcal{A}$ into several parallel classes is called a resolution, and $(X, \mathcal{A})$ is said to be a resolvable design if A has at least one resolution.
	\end{defn}
	\begin{exmp}
		\label{exmp1}
		Consider the block design specified as follows.
		\begin{align*}
		X =& \{1, 2, 3, 4\}, \text{ and}\\
		\mathcal{A} = &\{\{1, 2\},\{1, 3\},\{1, 4\},\{2, 3\},\{2, 4\},\{3, 4\}\}.
		\end{align*}
		It can be observed that this design is resolvable with the parallel classes 
		$\mathcal{P}_1 = \{\{1, 2\},\{3, 4\}\},$ $\mathcal{P}_2 = \{\{1, 3\},\{2, 4\}\},$  and $\mathcal{P}_3 = \{\{1, 4\},\{2, 3\}\}.$
		Note that $\mathcal{P}_1$, $\mathcal{P}_2$, and $\mathcal{P}_3$ form a partition of $\mathcal{A}$. If $\mathcal{A}$ = \{\{1, 2\},\{1, 3\},\{3, 4\},\{2, 4\}\}, we get another resolvable design with two parallel classes $\mathcal{P}_1$ and $\mathcal{P}_2$.
	\end{exmp}
	\begin{exmp}
		\label{exmp2}
		Consider the  block design specified as follows.
		\begin{align*}
		X =& \{1, 2, 3, 4, 5, 6\}, \text{ and}\\
		\mathcal{A} = &\{\{1, 2 ,3\},\{4, 5, 6\},\{1, 4 ,5\},\{2, 3, 6\}\}.
		\end{align*}
		It can be observed that this design is resolvable with two parallel classes: 
		$\mathcal{P}_1 =\{\{1, 2 ,3\},\{4, 5, 6\}\}$ and $\mathcal{P}_1 =\{\{1, 4 ,5\},\{2, 3, 6\}\}.$
	\end{exmp}
	
	For a given resolvable design  $(X, \mathcal{A})$ if |$X$| = $v$, |$\mathcal{A}$| = $b$ , block size is $k$ and the number of parallel classes is $r$, then there are exactly $\frac{b}{r}$ blocks in each parallel class. Since the blocks in each parallel class are disjoint, the number of blocks in each parallel class = $\frac{b}{r}$ = $\frac{v}{k}$.
\subsection{Cross Resolvable Design (CRD)\cite{KNRarXiv}}
	In \cite{KNRarXiv}, the notion of cross resolvable designs were introduced for the first time. In \cite{KNRarXiv}, multi-access coded caching schemes from CRDs were discussed and two classes of CRDs from affine geometry and Hadamard matrices were studied for use in multi-access systems.
\begin{defn}[\textbf{Cross Intersection Number}]
		For any resolvable design $(X, \mathcal{A})$ with $r$ parallel classes, the $i^{th}$ cross intersection number, $\mu_{i}$ where $i \in \{2, 3, \dots,r\}$, is defined as the cardinality of intersection of $i$ blocks drawn from any $i$ distinct parallel classes, provided that, this value remains same ($\mu_i\neq 0$), for all possible choices of blocks. \\	
		For instance, in Example 1, $\mu_{2}$ = 1 and  $\mu_{3}$ does not exist. 
	\end{defn}
\begin{defn}[\textbf{Cross Resolvable Design}]
		For any resolvable design $(X, \mathcal{A})$, if there exist at least one $i\in \{2,3,\dots,r\}$ such that the $i^{th}$ cross intersection number $\mu_{i}$ exists, then the resolvable design is said to be a Cross Resolvable Design (CRD). 
If $\mu_r$ exists then the CRD is said to be a Maximal Cross Resolvable Design (MCRD).
\end{defn}
Note that the resolvable design in \textit{Example 2} is not a CRD  as $\mu_{2}$ does not exist.
\begin{exmp}
		\label{exmp3}
		For the resolvable design $(X, \mathcal{A})$ with
		\begin{align*}
		X = &\;\{1, 2, 3, 4, 5, 6, 7, 8, 9\}, \text{ and} \\
		\mathcal{A} =& \;\{\{1, 2, 3\},\{4, 5, 6\},\{7, 8, 9\},\\
		&\{1, 4, 7\},\{2, 5, 8\},\{3, 6, 9\}\},
		\end{align*}
		the parallel classes are 
		$\mathcal{P}_1 = \;\{\{1, 2, 3\},\{4, 5, 6\},\{7, 8, 9\}\},$  and
		$\mathcal{P}_2 =\; \{\{1, 4, 7\},\{2, 5, 8\},\{3, 6, 9\}\}.$
		It is easy to verify that  $\mu_{2} = 1$.	
\end{exmp}
\begin{exmp}
		\label{exmp4}
		For the resolvable design $(X, \mathcal{A})$ with 
		\begin{align*}
		X =&\; \{1, 2, 3, 4, 5, 6, 7, 8\}, \text{ and} \\
		\mathcal{A} =&\; \{\{1, 2, 3, 4\},\{ 5, 6, 7, 8\},\{1, 2, 5, 6\},\\
		&\{3, 4, 7, 8\},\{1, 3, 5, 7\},\{2, 4, 6, 8\}\},
		\end{align*}
		the parallel classes are 
		$\mathcal{P}_1 =\; \{\{1, 2, 3, 4\},\{5, 6, 7, 8\}\},$ 
		$\mathcal{P}_2 =\; \{\{1, 2, 5, 6\},\{3, 4, 7, 8\}\},$  and 
		$\mathcal{P}_3 =\; \{\{1, 3, 5, 7\},\{2, 4, 6, 8\}\}.$
		We have $\mu_2=2$ and $\mu_3=1.$
\end{exmp}	
The multi-access coded caching scheme obtained from a CRD is as follows \cite{KNRarXiv}:  Consider a cross resolvable design $(X, \mathcal{A})$ with $v$ points, $r$ parallel classes, $b$ blocks of size $k$ each, $b_r\stackrel{def}{=}\frac{b}{r}$ blocks in each parallel class, the access degree $z \in \{2,3,\dots,r\}$ such that $\mu_{z}$ exists. A multi-access coded caching problem with $K = {r \choose z} (\frac{b}{r})^z$ number of users, $N$ files in server database, $b$  number of caches, $\frac{M}{N} = \frac{k}{v}$ fraction of each file at each cache and subpacketization level $v$ can associated to the CRD  $(X, \mathcal{A})$ \cite{KNRarXiv}. A user is connected to distinct $z$ caches such that these $z$ caches correspond to $z$ blocks from distinct parallel classes. The following theorem from \cite{KNRarXiv} gives the achievable rate obtained using the placement and delivery proposed in \cite{KNRarXiv}.
\begin{thm} \cite{KNRarXiv}
		\label{knrmathm}
		For $N$ files and $K$ users each with access to $z$
		caches of size $M$ in the considered caching system, if it $N \geq K$ and for the distinct demands by the users, the  scheme in \cite{KNRarXiv} achieves the rate $R$ given by
		$R = \frac{{\mu}_z {{b_r}\choose{2}}^z {{r}\choose{z}}}{v}$
	\end{thm}
\subsection*{Notations : }
	In this paper, $\mathbb{Z}_q$ denotes the set of integers modulo $q$, where $q$ is a positive integer, $\mathbb{Z}_q^t$ denotes the set of $t$ tuples over $\mathbb{Z}_q$, and $|\mathcal{X}|$ denotes the cardinality of a set $\mathcal{X}$.
\section{A NEW CLASS OF CRDs}
	Let $\mathcal{X} = \{0, 1, 2 \dots, q^m - 1\}$. 
	For $y \in \mathcal{X}$, let $\textbf{y} = (y_0, y_1, \dots, y_{m-1}) $ be the $q$-ary expansion of $\boldmath{y} = y_0 + y_1 q + y_2 q^2 + \dots + y_{m-1} q^{m-1} = \sum_{i = 0}^{m-1}y_i q^i$. For a positive integer $t$ such that $0<t<m$, we define a set of blocks as follows.\\
	Consider the set 
	\begin{align}
	\begin{split}
	 y_{a_{j_0}, a_{j_1}, a_{j_2},\dots,a_{j_{t-1}}} = \{\textbf{y} : y_{j_i} = a_{j_i} \forall\; 0\leq i<t\text{ where } \\
	 0 \leq j_0<j_1<j_2< \dots <j_{t-1}<m\}.
	\end{split}
	\end{align} 
	The set of blocks are
	\begin{align*}
	\begin{split}
	\mathcal{A} = \{  y_{a_{j_0}, a_{j_1}, a_{j_2},\dots,a_{j_{t-1}}}\; \forall\; \textbf{a} = (a_{j_0} a_{j_1} a_{j_2} \dots a_{j_{t-1}}) \in \mathbb{Z}_q^t\ :\\ 0 \leq j_0<j_1<j_2< \dots <j_{t-1}<m \}.
	\end{split}
	\end{align*} 
For every $\textbf{a} = (a_{j_0} a_{j_1} a_{j_2} \dots a_{j_{t-1}}) \in \mathbb{Z}_q^t$ with $0 \leq j_0<j_1<j_2< \dots <j_{t-1}<m$, there are ${m}\choose{t}$ ways of choosing such $\{j_0, j_1,\dots, j_{t-1}\}$, and for every such choice, there are $q^t$ ways of choosing $\textbf{a} \in \mathbb{Z}_q^t$. Thus there are totally $q^t {{m}\choose{t}}$ sets defined by (1). These $q^t {{m}\choose{t}}$ sets constitute the set of blocks in $\mathcal{A}$, leading to the design $(\mathcal{X}, \mathcal{A})$.
\begin{lem}
		The design constructed above is a resolvable design of parameters $v = q^{m}$, $b = q^{t} {{m}\choose{t}}$, $r = {{m}\choose{t}}, k = q^{m-t} $.	
\end{lem}
\begin{IEEEproof}
		It is easy to see that $v = |\mathcal{X}| = q^m	$ and $b = |\mathcal{A}| =  q^{t} {{m}\choose{t}}	$ by the construction. The block size $k$ is $q^{m-t}$ since for $\textbf{a} = (a_{j_0} a_{j_1} a_{j_2} \dots a_{j_{t-1}}) \in \mathbb{Z}_q^t$, the remaining $m - t$ positions of $\textbf{y}$ can be chosen from $\mathbb{Z}_q$ in $q^{m-t}$ ways.

Now we have to show that the design so obtained is resolvable.  Fix $j_0, j_1, j_2, \dots, j_{t-1}$ such that $0 \leq j_0 < j_1 < j_2 < \dots < j_{t-1} < m$. Consider $\textbf{a} = (a_{j_0}, a_{j_1}, a_{j_2}, \dots a_{j_{t-1}}) \in \mathbb{Z}_q^t$ and $\textbf{b} = (b_{j_0}, b_{j_1}, b_{j_2}, \dots b_{j_{t-1}}) \in \mathbb{Z}_q^t$.
		The intersection between any two blocks $y_{a_{j_0}, a_{j_1}, a_{j_2},\dots,a_{j_{t-1}}}$ and  $ y_{b_{j_0}, b_{j_1}, b_{j_2},\dots,b_{j_{t-1}}}$ is $\phi$. This can be proved as follows. If the intersection between any two such blocks is not empty then, for some \textbf{y}, $y_{j_i} = a_{j_i}$ and $y_{j_i} = b_{j_i}$ $\forall$  $0\leq i<t$. Then $\textbf{a} = \textbf{b}$ which is not possible, thus proving our claim.
		We have already proved that the size of the blocks are equal. Thus the set of blocks indexed by $a_{j_0} a_{j_1} a_{j_2} \dots a_{j_{t-1}}$ form a parallel class for a fixed choice of $j_0, j_1, j_2, \dots ,j_{t-1}$.
		There will be thus $q^t$ blocks in a parallel class. Since there are ${{m}\choose{t}}$ ways of choosing $j_0, j_1, j_2, \dots, j_{t-1}$ such that such that $0 \leq j_0 < j_1 < j_2 < \dots < j_{t-1} < m$ for a fixed choice of $\textbf{a}$, we have ${{m}\choose{t}}$ such parallel classes. 
	\end{IEEEproof} 
\begin{thm}
\label{crdthm1}
		When $t = 1$, the design from proposed construction gives a  cross resolvable design of parameters $v = q^{m}$, $b = q^{t} {{m}\choose{t}} = q m$, $k = q^{m-t} = q^{m-1} $, $r = {{m}\choose{t}} = m$ with cross intersection numbers $\mu_{z} = q^{m-z}$ for $z \in \{2,3,\dots,m\}$. The cross resolvable design thus obtained is a maximal cross resolvable design with $\mu_{m} = 1$.
\end{thm}
\begin{IEEEproof}
Consider any 2 blocks from 2 different parallel classes indexed by $a_{j_0}$ and $b_{k_0}$ for $0 \leq j_0,k_0 <m$. Clearly $j_0 \neq k_0$, as the blocks belong to different parallel classes. Consider the set of vectors, with each $\textbf{y}$ such that $y_{j_0} = a_{j_0} $ and  $y_{k_0} = b_{k_0}$. Now the vector $\textbf{y}$ can be chosen in $q^{m-2}$ different ways. Thus any two blocks drawn from two different parallel classes intersect at exactly $q^{m-2}$ points implying that $\mu_{2} = q^{m-2}$. If we consider a set of $z$ blocks drawn from $z$ distinct parallel classes, for instance $a_{j_0}^{(1)}, a_{j_0}^{(2)}, \dots , a_{_0}^{(z)}$, then the set of vectors with vector $\textbf{y}$ such that $y_{j_{0}}^{(1)} =a_{j_0}^{(1)}$, $y_{j_{0}}^{(2)} =a_{j_0}^{(2)}, \cdots y_{j_{0}}^{(z)} = a_{j_0}^{(z)}$ contains exactly  $q^{m-z}$ choices. Thus, the $z$th cross intersection number $\mu_{z} = q^{m-z}$.

Now if we consider a set of $m$ blocks drawn from $m$ distinct parallel classes, repeating same argument we have exactly one vector $\textbf{y}$ with ${y_{j_{0}}}^{(1)} = {a_{j_0}}^{(1)}$, ${y_{j_{0}}}^{(2)} ={a_{j_0}}^{(2)} $, $\dots$,  ${y_{j_{0}}}^{(m)} = {a_{j_0}}^{(m)}$. Thus  $\mu_{m} = 1$. Since $z = r$, in this case, the design formed by the construction is a maximal cross resolvable design.

\end{IEEEproof}
	\begin{exmp}
		In Table \ref{Examples}, for  the $\mathbb{Z}_{2^3}$ CRD it can be verified that $\mu_3 = 1$. For the $\mathbb{Z}_{2^4}$ CRD,  $\mu_{2} = 4$, $\mu_{3} = 2$ and $\mu_{4} = 1$. For the $\mathbb{Z}_{3^2}$ CRD we have 	$\mu_{2} = 1$ and for the $\mathbb{Z}_{4^2}$ CRD, $\mu_{2} = 1$.	
    \end{exmp}
\begin{table*}[h]
	\caption{Examples}
	\begin{center}
		\renewcommand{\arraystretch}{2}
		\begin{tabular}{|c|c|c|c|c|}
			\hline
			\textbf{Design} &\textbf{Parameters}  &  X & $\mathcal{A}$ &\textbf{Parallel classes}\\\hline\hline	
			$\mathbb{Z}_{2^3}$ CRD & $t = 1$, $m = 3$, $q = 2$ & 	$\{ 0, 1, 2, 3, 4, 5, 6, 7\}$ &  \makecell{$\{\{0, 1, 2, 3\},\{4, 5, 6, 7\},$\\
				$\{0, 2, 4, 6\},\{1, 3, 5, 7\},$\\
				$\{0, 1, 4, 5\},\{2, 3, 6, 7\}\}$ 
			}& \makecell{$\mathcal{P}_1 = \{\{\{0, 1, 2, 3\},\{4, 5, 6, 7\}\},$\\$\mathcal{P}_2 = \{ \{0, 2, 4, 6\},\{1, 3, 5, 7\}\},$\\$\mathcal{P}_3 = \{\{0, 1, 4, 5\},\{2, 3, 6, 7\}\}$}\\
			\hline
			$\mathbb{Z}_{2^3}$ RD & $t = 2$, $m = 3$, $q = 2$ & 	$\{ 0, 1, 2, 3, 4, 5, 6, 7\}$ &  \makecell{$\{\{0,1\},\{2,3\},\{4,5\},\{6,7\},$ \\
				$\{0,2\},\{1,3\},\{4,6\},\{5,7\},$ \\
				$\{0,4\},\{1,5\},\{2,6\},\{3,7\}\}$}& \makecell{ $\mathcal{P}_1 = \{\{0,1\},\{2,3\},\{4,5\},\{6,7\}\},$\\
				$\mathcal{P}_2 = \{ \{0,2\},\{1,3\},\{4,6\},\{5,7\}\},$\\ $\mathcal{P}_3 = \{\{0,4\},\{1,5\},\{2,6\},\{3,7\}\}$}\\
			\hline
			$\mathbb{Z}_{2^4}$ CRD & $t = 1$, $m = 4$, $q = 2$ & \makecell{	$\{ 0, 1, 2, 3, 4, 5, 6, 7,$\\
				$ 8, 9, 10, 11, 12, 13, 14, 15\}$} &  \makecell{$\{\{0,1,2,3,4,5,6,7\},$\\$\{8,9,10,11,12,13,14,15\},$\\
				$\{0,1,2,3,8,9,10,11\},$\\$\{4,5,6,7,12,13,14,15\},$\\
				$\{0,2,4,6,8,10,12,14\},$\\$\{1,3,5,7,9,11,13,15\},$\\
				$\{0,1,4,5,8,9,12,13\},$\\$\{2,3,6,7,10,11,14,15\}\}$}& \makecell{ $\mathcal{P}_1 = \{\{0,1,2,3,4,5,6,7\},$\\$\{8,9,10,11,12,13,14,15\}\},$\\
				$\mathcal{P}_2 = \{ \{0,1,2,3,8,9,10,11\},$\\$\{4,5,6,7,12,13,14,15\}\},$\\ $\mathcal{P}_3 = \{\{0,2,4,6,8,10,12,14\},$\\$\{1,3,5,7,9,11,13,15\}\},$\\	
				$\mathcal{P}_4 = \{\{0,1,4,5,8,9,12,13\},$\\$\{2,3,6,7,10,11,14,15\}\}$}\\
			\hline
			
			$\mathbb{Z}_{2^4}$ RD & $t = 2$, $m = 4$, $q = 2$ & \makecell{	$\{ 0, 1, 2, 3, 4, 5, 6, 7,$\\
				$ 8, 9, 10, 11, 12, 13, 14, 15\}$} &  \makecell{$\{\{0,1,2,3\}, \{4,5,6,7\},$\\$ \{8,9,10,11\}, \{12,13,14,15\},$\\
				$\{0,1,4,5\},\{2,3,6,7\},$\\$\{8,9,12,13\}, \{10,11,14,15\},$\\
				$\{0,2,4,6\}, \{1,3,5,7\},$\\$ \{8,10,12,14\},\{9,11,13,15\},$\\	
				$\{0,1,8,9\},\{2,3,10,11\},$\\$\{4,5,12,13\}, \{6,7,14,15\},$\\
				$\{0,2,8,10\},\{1,3,9,11\},$\\$\{4,6,12,14\}, \{5,7,13,15\},$\\
				$\{0,4,8,12\},\{1,5,9,13\},$\\$\{2,6,10,14\}, \{3,7,11,15\}\}$\\}& \makecell{  $\mathcal{P}_1 = \{\{0,1,2,3\}, \{4,5,6,7\},$\\$ \{8,9,10,11\}, \{12,13,14,15\}\},$\\
				$\mathcal{P}_2 = \{\{0,1,4,5\},\{2,3,6,7\} ,$\\$\{8,9,12,13\}, \{10,11,14,15\}\},$\\
				$\mathcal{P}_3 = \{\{0,2,4,6\}, \{1,3,5,7\},$\\$ \{8,10,12,14\},\{9,11,13,15\}\},$\\
				$\mathcal{P}_4 = \{\{0,1,8,9\},\{2,3,10,11\} ,$\\$\{4,5,12,13\}, \{6,7,14,15\}\},$\\
				$\mathcal{P}_5 = \{\{0,2,8,10\},\{1,3,9,11\} ,$\\$\{4,6,12,14\}, \{5,7,13,15\}\},$\\
				$\mathcal{P}_6 = \{\{0,4,8,12\},\{1,5,9,13\} ,$\\$\{2,6,10,14\}, \{3,7,11,15\}\}$}\\
			\hline
			
			$\mathbb{Z}_{2^4}$ RD & $t = 3$, $m = 4$, $q = 2$ & \makecell{	$\{ 0, 1, 2, 3, 4, 5, 6, 7,$\\
				$ 8, 9, 10, 11, 12, 13, 14, 15\}$} &  \makecell{$\{\{0,1\}, \{2,3\}, \{4,5\}, \{6,7 \},$\\$ \{8,9\}, \{10,11\}, \{12,13\}, \{14,15\},$\\
				$\{0,2\}, \{1,3\}, \{4,6\}, \{5,7\},$\\$ \{8,10\}, \{9,11\}, \{12,14\}, \{13,16\},$\\
				$\{0,4\}, \{1,5\}, \{2,6\}, \{3,7\},$\\$ \{8,12\}, \{9,13\}, \{10,14\}, \{11,15\},$\\
				$\{0,8\}, \{1,9\}, \{2,10\}, \{3,11\},$\\$ \{4,12\}, \{5,13\},\{6,14\}, \{7,15\}\}$}& \makecell{  $\mathcal{P}_1 = \{\{0,1\}, \{2,3\}, \{4,5\}, \{6,7 \},$\\$ \{8,9\}, \{10,11\}, \{12,13\}, \{14,15\}\},$\\
				$\mathcal{P}_2 = \{\{0,2\}, \{1,3\}, \{4,6\}, \{5,7\},$\\$ \{8,10\}, \{9,11\}, \{12,14\}, \{13,16\}\},$\\
				$\mathcal{P}_3 = \{\{0,4\}, \{1,5\}, \{2,6\}, \{3,7\},$\\$ \{8,12\}, \{9,13\}, \{10,14\}, \{11,15\}\},$\\
				$\mathcal{P}_4 = \{\{0,8\}, \{1,9\}, \{2,10\}, \{3,11\},$\\$ \{4,12\}, \{5,13\}, \{6,14\}, \{7,15\}\}$}\\
			\hline
			$\mathbb{Z}_{3^2}$ CRD & $t = 1$, $m = 2$, $q = 3$ & 	$\{ 0, 1, 2, 3, 4, 5, 6, 7, 8\}$ &  \makecell{$\{\{0, 1 ,2\},\{3, 4, 5\},\{6, 7, 8\},$\\$\{0, 3, 6\},\{1, 4, 7\},\{2, 5, 8\}\}$
			}& \makecell{$\mathcal{P}_1 = \{\{0, 1 ,2\},\{3, 4, 5\},\{6, 7, 8\}\},$\\$\mathcal{P}_2 = \{\{0, 3, 6\},\{1, 4, 7\},\{2, 5, 8\}\}\}$}\\
			\hline
			$\mathbb{Z}_{4^2}$ CRD & $t = 1$, $m = 2$, $q = 4$ & 	\makecell{$\{0, 1, 2, 3, 4, 5, 6, 7,$\\$ 8, 9, 10, 11, 12, 13, 14, 15\}$} &  \makecell{$\{\{0, 1, 2 ,3\},\{4, 5, 6, 7\},$\\$\{8, 9, 10, 11\},\{12, 13, 14, 15\},$\\
				$\{0, 4, 8, 12\},\{1, 5, 9, 13\},$\\$\{2, 6, 10, 14\},\{3, 7, 11, 15\}\}$
			}& \makecell{$\mathcal{P}_1 = \{\{0, 1, 2 ,3\},\{4, 5, 6, 7\},$\\$\{8, 9, 10, 11\},\{12, 13, 14, 15\}\},$\\$\mathcal{P}_2 = \{\{0, 4, 8, 12\},\{1, 5, 9, 13\},$\\$\{2, 6, 10, 14\},\{3, 7, 11, 15\}\}$}\\
			\hline
		\end{tabular}
	\end{center}
\label{Examples}
\end{table*}
	\begin{thm}
		\label{mathm1}
		The cross resolvable design in Theorem \ref{crdthm1} gives a coded caching scheme for multiaccess network with parameters $K = q^{z} {{m}\choose{z}}$, $\frac{M}{N} = \frac{1}{q}$, $F = q^{m}$, $R = {{m}\choose{z}} \frac{{(q - 1)}^z}{{2}^z}$ for $z \in \{2,3,\dots,m\}$. 
	\end{thm}
	\begin{IEEEproof}
		In 	Theorem \ref{knrmathm}, if the parameters of the constructed cross resolvable design are substituted, the result is obtained.
	\end{IEEEproof}

	\begin{rem}
		The choice $z = m$, gives the best per user rate.\\
		When we compare the multi access networks with access degrees $z$ and $z-1$, for $z \in \{3,4,\dots,m\}$
		\begin{align*}
		\frac{K_z}{K_{z-1}} &=\frac{ q^{z} {{m}\choose{z}}}{q^{z-1} {{m}\choose{z-1}}} \\&= q \Big(\frac{m - z + 1}{z}\Big)\\
		\frac{R_z}{R_{z-1}} &= \frac{{{m}\choose{z}} \frac{{(q - 1)}^z}{{2}^z}}{{{m}\choose{z-1}} \frac{{(q - 1)}^{z-1}}{{2}^{z-1}}} \\&= \Big(\frac{q-1}{2}\Big)\Big(\frac{m - z + 1}{z}\Big) \\
		{\frac{\frac{R_z}{K_z}}{\frac{R_{z-1}}{K_{z-1}}}} &= \frac{q-1}{2q}
		\end{align*}
	\end{rem}
	\begin{cor}
\label{cor1}
		The cross resolvable design in Theorem \ref{crdthm1} gives a coded caching scheme for multiaccess network with parameters $K = q^{m}$, $\frac{M}{N} = \frac{1}{q}$, $F = q^{m}$, $R = \frac{{(q - 1)}^m}{{2}^m}$ for $z = m$.
	\end{cor}
	\begin{rem}
		The multi-access scheme obtained from this construction has subpacketization level $F$ equal to the number of users $K$ when $z=m.$ Hence subpacketization $F$ grows linearly with $K$.
	\end{rem}
	\begin{rem}
		An interesting point to note in Corollary  \ref{cor1} is that the rate $R$ decreases as $m$ increases only if $q = 2$. However the per user rate $\frac{R}{K} = {\big(\frac{q - 1}{2q}\big)}^m$ decreases as $m$ increases for any value of $q$.
	\end{rem}
\begin{lem}
		Let $M'$ denotes the size of all the distinct subfiles accessible to a user after accessing multiple caches and access degree be $z$. Then we have
		$$\frac{M'}{N} = 1 - \Big(1 - \frac{1}{q}\Big)^z$$
		for the multi-access coded caching scheme in Theorem \ref{mathm1}.
\end{lem}
\begin{IEEEproof}
In \cite{KNRarXiv}, it is shown that $$\frac{M'}{N}= \frac{zM}{N}+\sum_{t=2}^z (-1)^{t+1}\binom{z}{t}\frac{\mu_t}{v}.$$\\
			Taking $\frac{M}{N} = \frac{1}{q}$, $\mu_t = q^{m-t}$ we get, \\
			\begin{equation*}
			\frac{M'}{N} = \frac{z}{q} + \sum_{t=2}^z (-1)^{t+1}\binom{z}{t}\frac{q^{m-t}}{q^m}
			= 1 - \Big(1 - \frac{1}{q}\Big)^z
			\end{equation*}
			where the last step follows from binomial expansion.	
		\end{IEEEproof}
\begin{lem}
\label{onlymaxcrdlem}
Suppose there exists a maximal CRD of parameters $v,b,r,\mu_{r} = 1$ then $v = b_r ^ r$.
\end{lem}
\begin{IEEEproof}
Any $r$ blocks drawn from $r$ distinct parallel classes intersect at a point. There are totally $b_r ^ r$ ways of doing this and hence $b_r ^ r$ distinct points. So $v \geq b_r ^ r$. 
Now to show that $v = b_r ^ r$, consider a point in a block. Due to the resolvability of the design, this point has to be present in some block in every parallel class. Intersection of all such blocks would then give this unique point, since the design has $\mu_r = 1$. So $v = b_r ^ r$.
\end{IEEEproof}
\section{Performance Analysis}
We refer to the construction of CRD in Theorem \ref{crdthm1} as the proposed construction henceforth and coded caching scheme obtained in Theorem \ref{mathm1} as the proposed scheme. In subsection \ref{CompMaN}, we compare the parameters of scheme derived from proposed construction with the MaN scheme. In subsections \ref{Compknr}, \ref{CompMultiaccess} we compare the multi-access scheme from the proposed construction to the existing multi-access schemes.
\subsection{Comparison of the proposed scheme with the MaN scheme}
	\label{CompMaN}
	We compare the proposed scheme with the MaN scheme keeping the number of caches and memory fraction $\frac{M}{N}$ same as shown in  Table \ref{compwithMaN}.   
\begin{table}[]
		\caption{Comparison between the MaN\cite{MaN} and the proposed scheme}
		\begin{center}
			\renewcommand{\arraystretch}{2}
			\begin{tabular}{|c|c|c|}
				\hline
				\textbf{Parameters} &\textbf{MaN Scheme}  &  \textbf{Proposed Scheme}\\\hline\hline
				Number of Caches & $qm$ &  $qm$\\\hline
				\makecell{File fraction $\left(\frac{M}{N}\right)$} & $\frac{1}{q}$ &  $\frac{1}{q}$\\\hline
				Number of Users $(K)$ & $qm$   &$q^z{{m}\choose{z}}$\\\hline
				Access Degree $(z)$ &  $1$ & $z$ \\\hline
				Subpacketization level  & $\binom{qm}{m}$ & $q^m$\\[.2cm]\hline
				Rate $(R)$  & $\frac{(q-1)m}{1+m}$ & $ {{m}\choose{z}} \frac{{(q - 1)}^z}{{2}^z}$\\[.2cm]\hline
				Rate per user $\left(\frac{R}{K}\right)$  & $\frac{q-1}{q(1+m)}$ & $({\frac{q-1}{2q}})^z$\\[.2cm]\hline
				Gain $(g)$ & ${m + 1}$  & $2^z$\\\hline
			\end{tabular}
		\end{center}
\label{compwithMaN}
\end{table}
	It is seen that the proposed construction allows the support of a large number of users while the rate per user decreases. If we fix $q = 2$, then it is seen that the rate of the MaN scheme is $\frac{m}{m+1}$ while the rate of the proposed scheme is $\binom{m}{z}0.5^z$.
	
\begin{figure}
    	\begin{center}
    		\includegraphics[width=9cm,height=9cm]{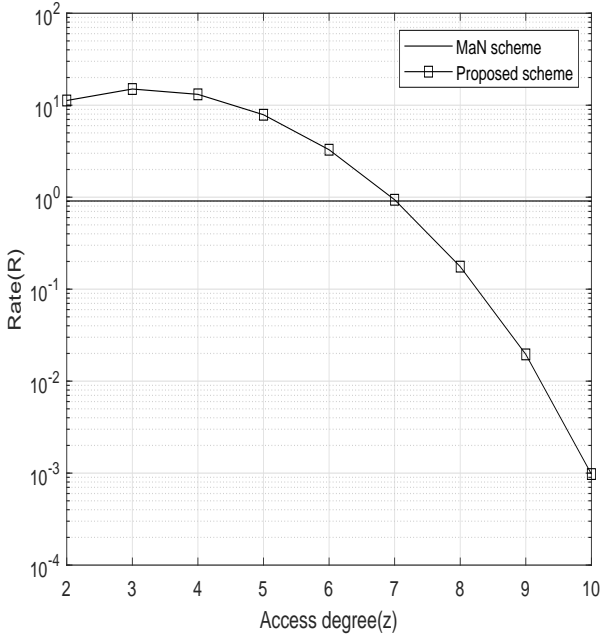}
    		\caption {Comparison of rates of the MaN scheme and scheme from proposed CRD for different values of access degree $z$ and $m = 10, q = 2$.
    		}
    		\label{Comprateman}
    	\end{center}
        \end{figure}

	When $\frac{m}{m+1} > \binom{m}{z} \frac{1}{2^z}$ and $q = 2,$  lower rates than the MaN scheme is achieved. When $z = m$, we have $\frac{m}{m+1} > 0.5^m$, and if $q = 2$, then lower rates than the MaN scheme is achieved for all choices of $m$. When $z = m-1$, for $m = 3$ and $q = 2$ the rates become the same, while for $m > 3$, lower rates than the MaN scheme is achieved. 
	Figure \ref{Comprateman} shows the comparison of rates of the MaN scheme and the proposed scheme for different values of access degree $z$ and $m = 10$, $q = 2$. The per user rate is less for the proposed scheme irrespective of the value of $q$ if $2^z > m + 1$. The proposed scheme has lesser subpacketization level than that of the MaN scheme.

    Now we calculate the access gain.
    \begin{align*}
   g_{access} &= \frac{\splitfrac{\text{Per user rate of optimal centralized coded caching}}{\text{scheme  with dedicated cache}}}{\text{Per user rate of multi-access system}}\\
     &= \frac{\frac{q-1}{q(1+m)}}{({\frac{q-1}{2q}})^z}\\
     &= {\Big(\frac{q}{q-1}\Big)}^{z - 1} \frac{2^z}{1+m}
    \end{align*}
    It is seen that for $z = m$, the access gain is maximum for some choice of $q$. Over all choices of $q$ for the choice $q = 2$ and $z = m$, the access gain becomes maximum.\\

\subsection{Comparison of the proposed scheme with  schemes in \cite{KNRarXiv}}
 \label{Compknr}
   \subsubsection{Comparison of the proposed scheme with those derived from CRDs from affine geometry in \cite{KNRarXiv}}
   \label{Compaffinegeo}
	A comparison of the multi-access scheme from the proposed  construction with the scheme derived from affine geometry is made in Table \ref{compwithaffinegeo}. The affine geometry parameters are $q$, where $q$ is a prime or a prime power and $m'$.
	To compare the proposed scheme with the scheme from affine geometry, the number of caches and memory fraction $\frac{M}{N}$ is kept the same.
	The scheme from affine geometry only exists when the number of blocks in the CRD is $\frac{q(q^{m'}-1)}{q-1}$ where $q$ is a prime or a prime power and $m' \geq 2$. Therefore to compare, we choose a CRD from the proposed construction with  $m = \frac{(q^{m'}-1)}{q-1}$ parallel classes. 
	Note that the access degree of the proposed scheme is kept maximum in Table \ref{compwithaffinegeo} for obtaining the best per user rate.
	\begin{table*}[]
		\caption{Comparison between the proposed scheme and scheme derived from CRDs from affine geometry}
		\begin{center}
			\renewcommand{\arraystretch}{2}
			\begin{tabular}{|c|c|c|}
				\hline
				\textbf{Parameters} &\textbf{Scheme in \cite{KNRarXiv}} & \textbf{Proposed Scheme}\\\hline\hline
				\makecell{Number of Caches} & $\dfrac{q(q^{m'}-1)}{q-1}$ & $\dfrac{q(q^{m'}-1)}{q-1}$\\[.2cm]\hline
				\makecell{File Fraction $\left(\frac{M}{N}\right)$} & $\dfrac{1}{q}$ & $\dfrac{1}{q}$\\\hline
				\makecell{Number of Users $(K)$} & $\dfrac{q^3(q^{m'}-1)(q^{{m'}-1}-1)}{2(q-1)^2}$ & $q^{\frac{q^{m'}-1}{q-1}}$ \\[.2cm]\hline
				Access Degree $(z)$ &  $2$ & $\dfrac{q^{m'}-1}{q-1}$  \\\hline
				\makecell{Subpacketization $(F)$} &  $q^{m'}$ & $q^{\frac{q^{m'}-1}{q-1}}$ \\[.2cm]\hline
				Rate $(R)$ & $\dfrac{q(q^{m'}-1)(q^{{m'}-1}-1)}{8}$ & $\Big(\frac{q-1}{2}\Big)^{\frac{q^{m'}-1}{q-1}}$  \\[.2cm]\hline
				\makecell{Rate per\ user $\left(\frac{R}{K}\right)$}&$\Big(\frac{q-1}{2q}\Big)^{2}$ & $\Big(\frac{q-1}{2q}\Big)^{\frac{q^{m'}-1}{q-1}}$\\[.2cm]\hline
				Gain $(g)$ & $2^2$ & $2^{\dfrac{q^{m'}-1}{q-1}}$  \\[.2cm]\hline
			\end{tabular}
		\end{center}
			\label{compwithaffinegeo}
	\end{table*}
	For any value of $q$, it is seen that the per user rate is smaller in the case of the proposed construction. However it is seen that this is happening by trading off at subpacketization levels. The subpacketization levels of the proposed scheme is much larger at large values of $q$ when compared to the scheme derived from affine geometry.

\begin{figure}
		\begin{center}
			\includegraphics[width=9cm,height=9cm]{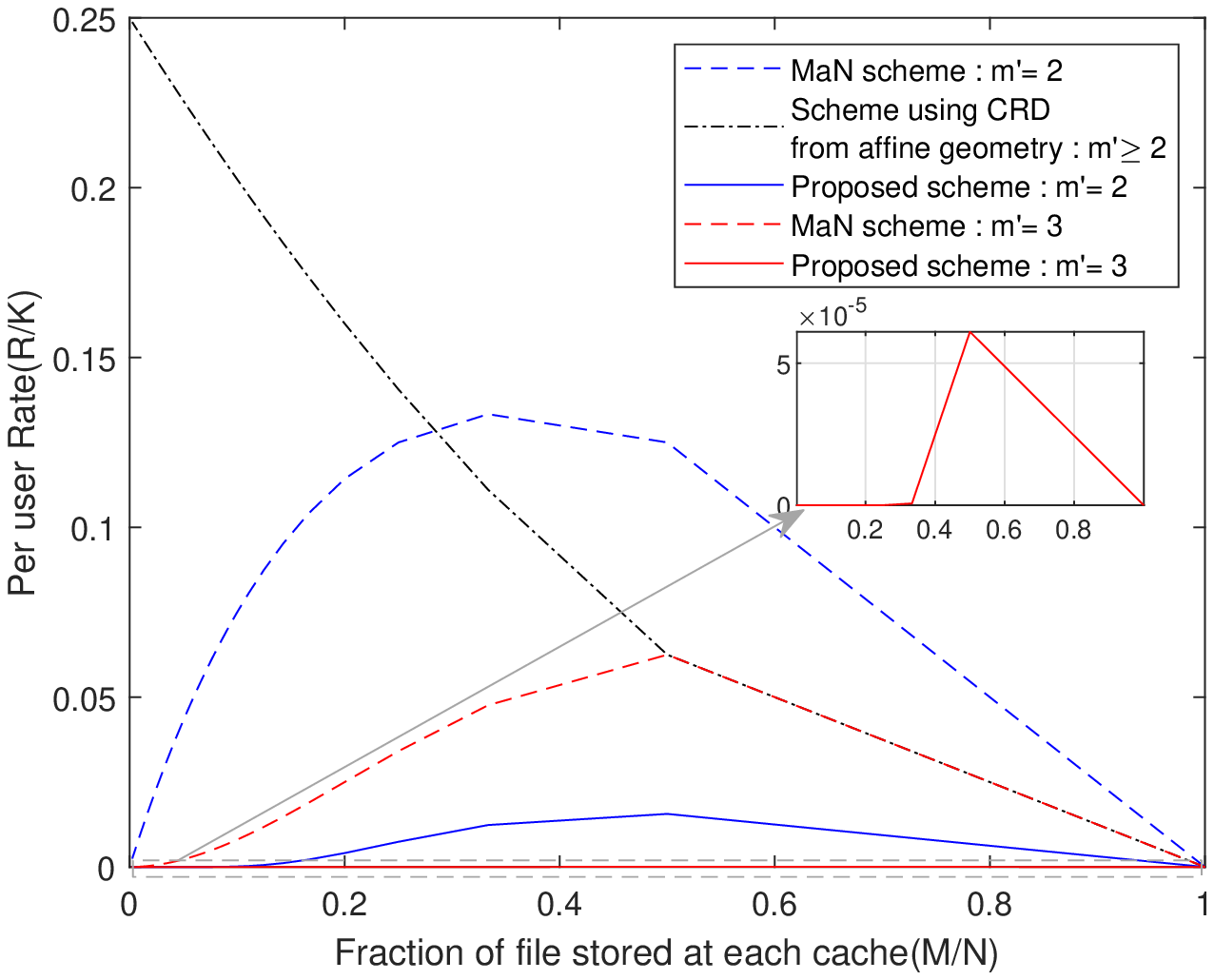}
			\caption {Comparison of per-user rates of the MaN scheme, the scheme for the
				class of CRD derived from affine geometry and the proposed scheme for different values of $m’$.
			}
			\label{ComprateperuserMan}
		\end{center}
\end{figure}

\begin{figure}
		\begin{center}
			\includegraphics[width=10cm,height=9cm]{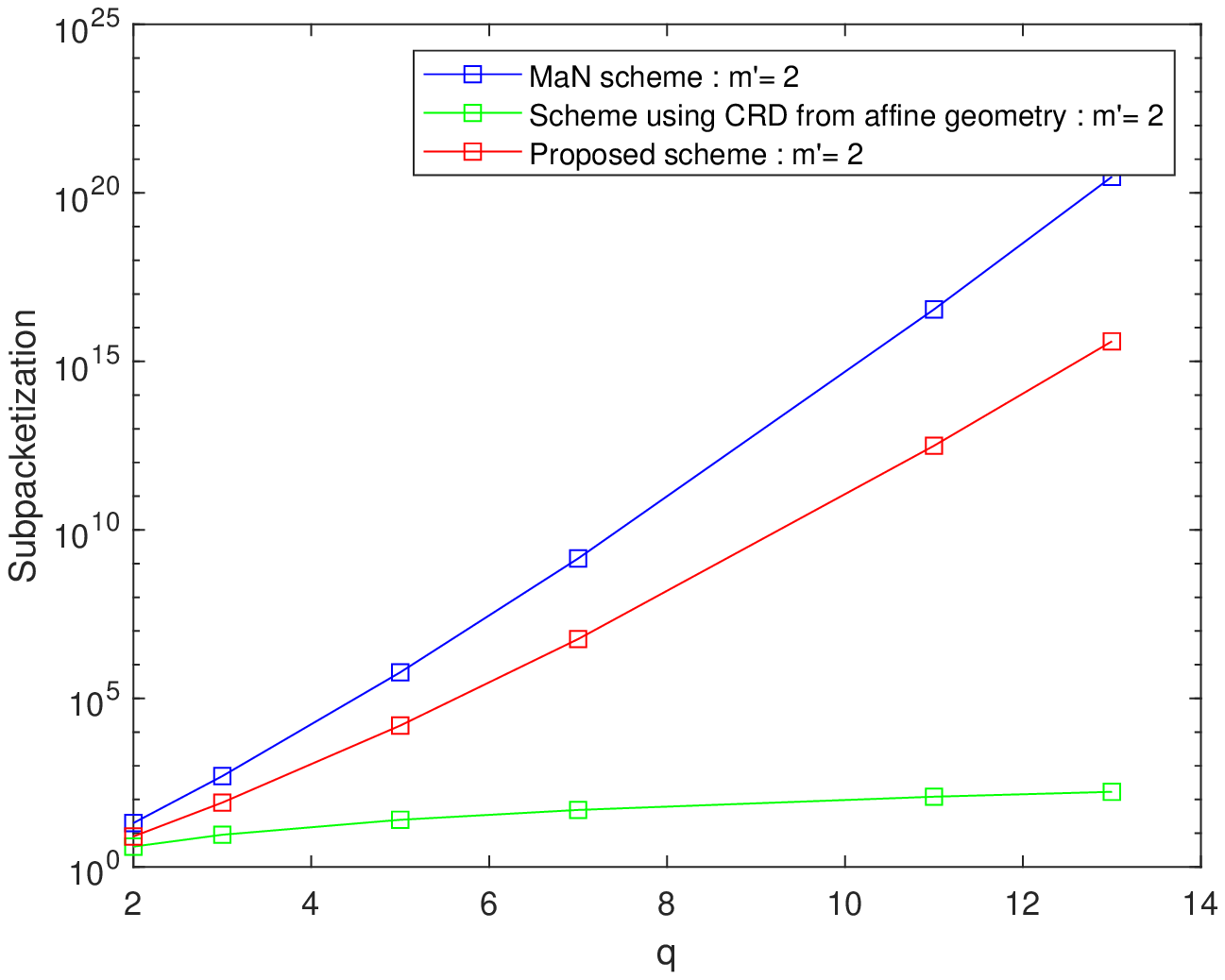}
			\caption {Comparison of subpacketization levels of the MaN scheme, the scheme for the class of CRD derived from affine geometry and the proposed scheme for $m’ = 2$.	
			}
			\label{CompsubpacketMan}
		\end{center}
\end{figure}

	Figure \ref{ComprateperuserMan} shows the comparison of per-user rates of MaN scheme, scheme for the
	class of CRD derived from affine geometry and scheme from proposed CRD for different values of $m’$.
    Figure \ref{CompsubpacketMan} shows the comparison of subpacketization levels of MaN scheme, scheme for the
    class of CRD derived from affine geometry and scheme from proposed CRD for $m’ = 2$.
\subsubsection{Comparison of the proposed scheme with those derived from CRDs from Hadamard matrices in \cite{KNRarXiv}}
	\label{CompHadamard}
	The  proposed scheme is compared with the scheme in \cite{KNRarXiv} derived from CRDs from Hadamard matrices in Table \ref{compwithhadamard}. 
	The number of caches and memory fraction $\frac{M}{N}$ is kept same. Note that the access degree of proposed scheme is kept maximum in Table \ref{compwithhadamard} for obtaining the best per user rate.
	{\small 
		\begin{table*}[]
			\caption{Comparison between Proposed scheme and scheme for the class of cross resolvable design derived from Hadamard matrices\cite{KNRarXiv}}
				
			\begin{center}
				\renewcommand{\arraystretch}{2.5}
				\begin{tabular}{|c|c|c|}
					\hline
					\textbf{Parameters} &\textbf{Scheme in \cite{KNRarXiv} } & \textbf{Proposed Scheme }\\\hline\hline
					\makecell{Number of Caches} & $2(4n-1)$ & $2(4n-1)$\\\hline
					\makecell{File fraction $\left(\frac{M}{N}\right)$} &$\frac{1}{2}$ & $\frac{1}{2}$  \\\hline
					\makecell{Number of Users $(K)$} & $4(2n-1)(4n-1)$ & $2^{4n-1}$ \\\hline
					Access Degree $(z)$ & $2$ & $ $4n-1$$ \\\hline
					\makecell{Subpackelization level $(F)$} & $4n$ & $2^{4n-1}$  \\\hline
					Rate $(R)$ & $\frac{(2n-1)(4n-1)}{4}$ & $\frac{1}{2^{4n -1}}$   \\\hline
					Rate per user $\left(\frac{R}{K}\right)$ &$\frac{1}{16}$ &$\frac{1}{4^{4n -1}}$\\\hline
					Gain $(g)$ & $4$ & $2^{4n}$ \\\hline
				\end{tabular}
			\end{center}
			\label{compwithhadamard}
		\end{table*}
	}

	The subpacketization levels of the proposed scheme is much larger at large values of $q$ when compared to the scheme derived from Hadamard matrices. But it can be observed that the number of users is also larger for the proposed scheme at the same cost, i.e., the number of caches and fraction of files stored in the cache being the same. The rate and rate per user are also lower in the multi-access scheme given by the proposed construction.
\subsection{Comparison of the proposed scheme with other multi-access schemes}
		\label{CompMultiaccess}
		In this subsection, we compare the performance of the proposed scheme with other multi-access schemes. 
\subsubsection{Comparison with the SPE scheme \cite{SPE}}
\begin{figure}
        	\includegraphics[width=10cm,height=9cm]{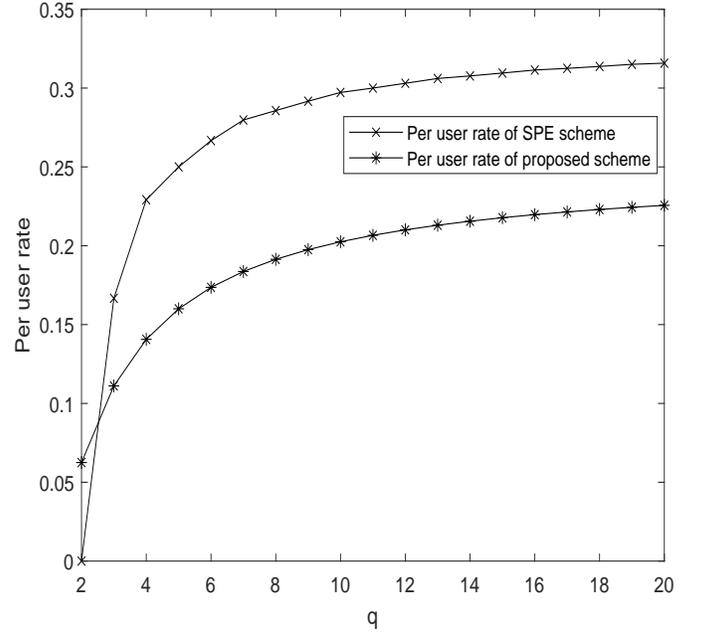}
        	\begin{center}
        		\caption {Comparison of per user rates of the proposed scheme vs the SPE scheme for different values of $q$.}
        		\label{fig2}
        	\end{center}
        \end{figure}

		For a fair comparison we keep the number of caches $C$, $\frac{M}{N}$ and access degree $z$, same for both schemes. In the SPE scheme, $C\frac{M}{N} = 2$.
		To compare with our scheme, thus we take $m = 2$ and hence, $C = 2q$.
		Since $z \leq m$, for the proposed scheme, in order to compare we fix $z = 2$.
		For the SPE scheme,
		$F =\frac{C(C - 2z + 2)}{4}$. On substituting,
		we get, $F =\frac{C(C - 2z + 2)}{4} = \frac{2q(2q - 2)}{4} = q(q-1)$. For the proposed scheme, subpacketization $F = q^2$.
		We see that subpacketization level of the SPE scheme is slightly less than that of proposed scheme.
		The rate per user of both the schemes is then compared. Figure \ref{fig2} shows the comparison of per user rates of the proposed scheme vs the SPE scheme for different values of $q$.\\	
          We see that except when $q = 2$, our scheme performs better than the SPE scheme, while sacrificing slightly in terms of subpacketization.
\subsubsection{Comparison with the RK scheme \cite{RaK3}}
		An achievable scheme in \cite{RaK3}, gives rate $R_{RK} = C \big(1 - z \frac{M}{N}\big)^2$. To compare the proposed scheme with the RK scheme, we keep the number of caches as $qm$, access degree $z$, and $\frac{M}{N}$ as $\frac{1}{q}$.
		Then $$\Big(\frac{R}{K}\Big)_{RK} =  \Big(1 -  \frac{z}{q}\Big)^2$$ For the special case of $z > \frac{C}{2}$, the authors give a lower bound which is achievable for certain choices of $z$. 
		For the performance comparison in this section, we use the rate given by the achievable scheme which holds for any value of $z$.
		Rate of this scheme goes to zero when $\frac{zM}{N} = 1$.
		To allow a fair comparison of both the schemes, we compare both the schemes in terms of per-user rates.
		\begin{align*}
		\Big(\frac{R}{K}\Big)_{prop} = \Big({\frac{1 - \frac{1}{q}}{2}}\Big)^z \;\text{for}\; z \in \{2,3,\dots,m\}
		\end{align*}
		
\begin{figure}
			\includegraphics[width=10cm,height=9cm]{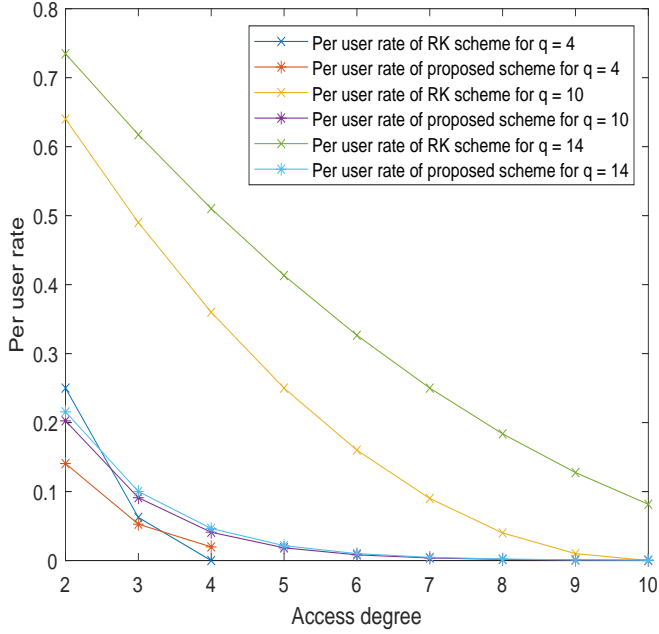}
			\begin{center}
				\caption {Comparison of per user rate of the proposed scheme vs the RK scheme for different values of $z$, $q$ with $m = 10$.}
				\label{fig3}
			\end{center}
\end{figure}

\begin{figure}
			\includegraphics[width=10cm,height=9cm]{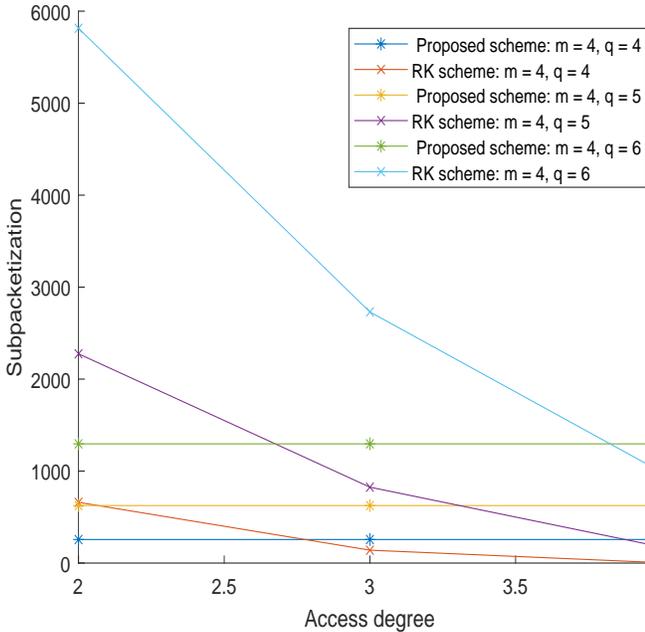}
			\begin{center}
				\caption {Comparison of subpacketization levels of the proposed scheme vs the RK scheme for different values of $z$, $q$ with $m$.}
				\label{fig4}
			\end{center}
\end{figure}

		Figure \ref{fig3} shows the comparison of per user rates of the proposed scheme vs the RK scheme for different values of $z$, $q$ with $m = 10$.
Subpacketization is $ \binom{C - iz + i-1}{i-1} \frac{C}{i}$, where $i = \frac{CM}{N}$. Taking $C = qm$, $\frac{M}{N}= \frac{1}{q}$, subpacketization becomes $ \binom{qm - mz + m-1}{m-1} q$. 	Subpacketization levels are independent of access degree for CRD based schemes, while it is dependent on access degree for the RK scheme. For small access degrees, the proposed scheme is better than the RK scheme in terms of subpacketization. Figure \ref{fig4} shows the comparison of subpacketization levels of proposed scheme vs RK scheme for different values of $z$, $q$ and $m$.
\subsubsection{Comparison with the CLWZC scheme  \cite{CLWZC} }
In this section, we  compare the results with the work in \cite{CLWZC}. First we study the variation of $\frac{R}{K}$ with respect to $\frac{M}{N}$ for both the schemes with $\frac{M}{N}$, kept constant. 
		We keep $\frac{M}{N} = \frac{1}{q}$, number of caches = $mq$ and access degree $z$ same for both schemes. The number of users will be $qm$ for the scheme in \cite{CLWZC} and the number of users will be $q^z \binom{m}{z}$ for the proposed scheme. It can be seen that the number of users supported in the proposed scheme is exponentially larger than the scheme in \cite{CLWZC} due to the difference in the setup. 
		For the scheme in \cite{CLWZC},\\
		\begin{align*}
		\Big(\frac{R}{K}\Big)_{CLWZC} &= \frac{qm - \frac{zqm}{q}}{qm(1 + \frac{qm}{q})} = \frac{1 - \frac{z}{q}}{1 + m}
		\end{align*}
		
		The rate in \cite{CLWZC} goes to zero when $z \geq q$. 
		But this is not the case with the CRD based multi-access schemes.
		It can be seen that the parameters of the chosen CRD are playing a role here. 
		 Another observation is that that $\big(\frac{R}{K}\big)_{prop}$ drops faster to zero with respect to access degree $z$ when compared to $\big(\frac{R}{K}\big)_{CLWZC}$ which drops linearly with respect to $z$. The performance of per user rate of the proposed scheme may or may not be better than the scheme in \cite{CLWZC}, based on the choices of $q$, $m$ and $z$.	
		\begin{align*}
		 \Big(\frac{R}{K}\Big)_{prop} &=\Big({\frac{1 - \frac{1}{q}}{2}}\Big)^z \;\text{for}\; z \in \{2,3,\dots,m\}\\
		 &\geq \frac{1 - \frac{z}{q}}{2^z}\; \text{for}\;z<q \;\text{and}\;  z \in \{2,3,\dots,m\} 
		\end{align*}
		where the last step follows from Bernouilli's inequality($(1 - x)^t \geq 1-xt, 0 \leq x \leq 1, t \geq 1$).
		For $z << q$, and $2^z > m + 1$ or $z < q$ and $2^z >> m + 1$, the per user rate of proposed scheme is better than that of the scheme of \cite{CLWZC}.
\begin{figure}
			\includegraphics[width=10cm,height=9cm]{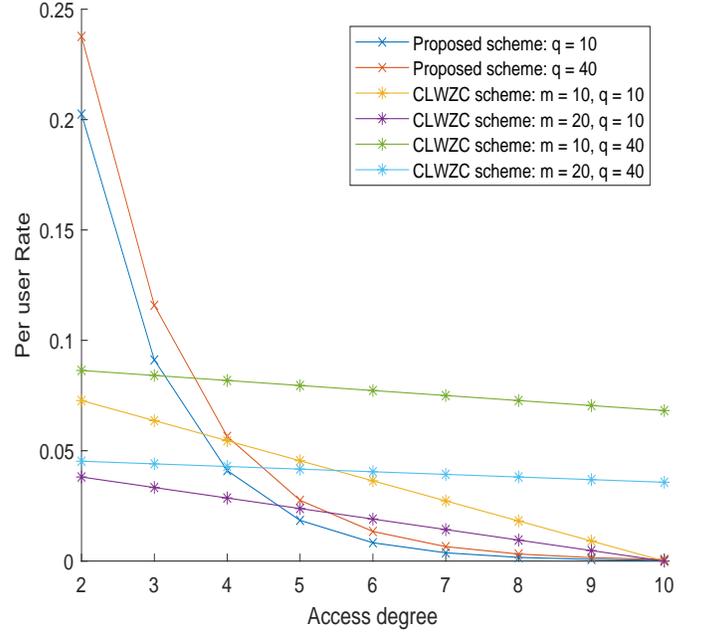}
			\begin{center}
				\caption {Comparison of per user rates of the proposed scheme vs the CLWZC scheme for $z \in \{2,3,\dots,10\}$}
				\label{fig5}
			\end{center}
\end{figure} 
 Figure \ref{fig5} shows the comparison of per user rates of the proposed scheme vs the CLWZC scheme.
%
%
%
%
	
\begin{figure}
	\includegraphics[width=10cm,height=9cm]{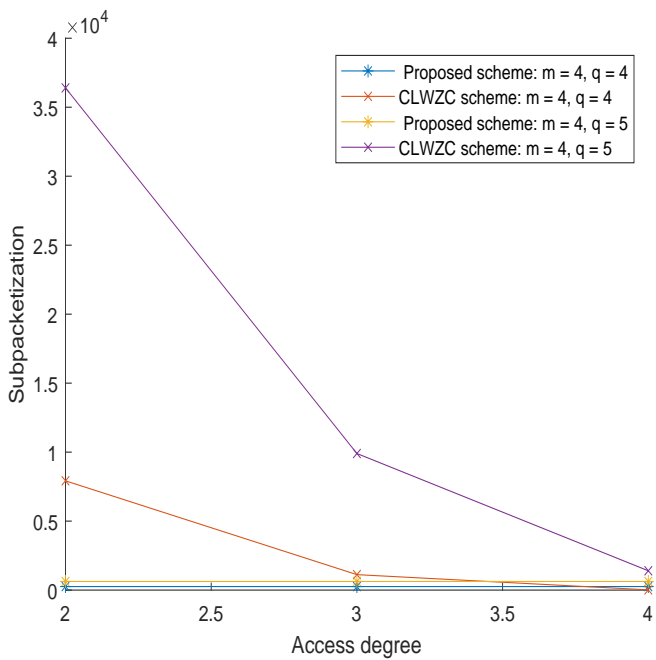}
	\begin{center}
		\caption {Comparison of subpacketization levels of the proposed scheme vs the CLWZC scheme for different values of $z$, $q$ and $m$}
		\label{fig6}
	\end{center}
\end{figure}

In terms of subpacketization, the scheme from proposed construction has the subpacketization level $q^m$ as opposed to $ qm\binom{qm - m(z-1)}{m}$ for the scheme in \cite{CLWZC}.
Subpacketization levels are independent of access degree for CRD based schemes, while it is dependent on access degree for the CLWZC scheme. For small access degrees, the proposed scheme is better than the CLWZC scheme in terms of subpacketization. Figure \ref{fig6} shows the comparison of subpacketization levels of the proposed scheme vs the CLWZC scheme for different values of $z$, $q$ and $m$.
For large values of $m$ and $q = 2$, it can be seen that for $z = m$, the proposed scheme has a rate close to $0$, while rate of scheme proposed in \cite{CLWZC} is $0$. But it is interesting to see that in this case, we can support $2^m$ users in proposed scheme as opposed to only $2m$ users in \cite{CLWZC}. An exponentially large number of users can be hence supported in the proposed scheme using CRD with the same storage.

	\section{Conclusion}
	The proposed construction of the  CRDs and the corresponding  multi-access schemes allow for a wide range of parameters as the parameter $q$, and parameter $m$ to be chosen for the construction can be any integer greater than or equal to 2.  The multi-access coded caching schemes presented in \cite{KNRarXiv} give smaller per user rate than the MaN scheme only in the high cache storage regime whereas the CRDs presented in this paper give smaller per user rate for all cache storage regime. It is also found to provide lower per-user rates than all other CRDs known in literature. In some cases, even the rates obtained are lower than the rates derived from multi-access schemes from other CRDs. It is also seen that based on the choice of parameters, the new construction can support a larger number of users in the multi-access network when the memory fraction stored in the cache and the number of caches are kept the same as compared to other known constructions. The construction for CRDs proposed provide a lot of flexibility in design and advantages, mainly with respect to the rates and rate per user metric.	

	\section*{Acknowledgment}
	This work was supported partly by the Science and Engineering Research Board (SERB) of Department of Science and Technology (DST), Government of India, through J.C. Bose National Fellowship to B. Sundar Rajan.

	

\begin{thebibliography}{1}
		
		\bibitem{MaN}
		M.~A. Maddah-Ali and U.~Niesen, ``Fundamental limits of caching,'' \emph{{IEEE}
			Trans. Inf. Theory}, vol.~60, no.~5, pp. 2856--2867, May 2014.
		\bibitem{SBP2}
		E. Parrinello, A. Unsal, and P. Elia, ``Coded caching with shared caches: Fundamental limits with uncoded prefetching,”
		\textit{arXiv preprint arXiv:1809.09422}, 2018
		\bibitem{AAA}
		A. M. Ibrahim, A. A. Zewail and A. Yener, "Coded Placement for Systems with Shared Caches," ICC 2019 - 2019 IEEE International Conference on Communications (ICC), Shanghai, China, 2019, pp. 1-6.
		\bibitem{HKD}
		J.Hachem, N. Karamchandani, and S. N. Diggavi, ``Coded caching for multi-level popularity and access," in \textit{IEEE Transactions on Information Theory}, vol. 63, no. 5, pp. 3108–3141, 2017
		\bibitem{SPE}
		Berksan Serbetci, Emanuele Parrinello and Petros Elia, ``Multi-access coded caching: gains beyond cache-redundancy” \textit{IEEE Information Theory Workshop}, Visby, Gotland, 2019
		\bibitem{RaK3}
		K. S. Reddy and N. Karamchandani, ``Rate-Memory Trade-off for Multi-Access Coded Caching With Uncoded Placement," in IEEE Transactions on Communications, vol. 68, no. 6, pp. 3261-3274, June 2020
		\bibitem{KNRarXiv} Digvijay Katyal, Pooja Nayak M. and B. Sundar Rajan,  "Multi-access Coded Caching Schemes From Cross Resolvable Designs," IEEE Transactions on Communications, (Accepted for publication)(Available as early access article in IEEE Xplore on \url{https://ieeexplore.ieee.org/stamp/stamp.jsp?tp=&arnumber=9328826})
		\bibitem{CLWZC}
		Minquan Cheng, Dequan Liang, Kai Wan, Mingming Zhang, Giuseppe Caire
		``A Novel Transformation Approach of Shared-link Coded Caching Schemes for Multiaccess Networks," Available on  arXiv:2012.04483 [cs.IT].
		\bibitem{BL}
		B. Asadi and L. Ong, "Centralized Caching with Shared Caches in Heterogeneous Cellular Networks," 2019 IEEE 20th International Workshop on Signal Processing Advances in Wireless Communications (SPAWC), Cannes, France, 2019, pp. 1-5.
		\bibitem{TaR}
		L. Tang and A. Ramamoorthy, ``Coded Caching Schemes With Reduced Subpacketization From Linear Block Codes," in \textit{IEEE Transactions on Information Theory}, vol. 64, no. 4, pp. 3099-3120, April 2018.
	
		\bibitem{RaK}
		K. S. Reddy and N. Karamchandani, ``On the Exact Rate-Memory Trade-off for Multi-access Coded Caching with Uncoded Placement," \textit{2018 International Conference on Signal Processing and Communications (SPCOM)}, Bangalore, India, 2018, pp. 1-5.
		
		\bibitem{RaK2}
		K. S. Reddy and N. Karamchandani, ``Rate-memory trade-off for multi-access coded caching with uncoded placement,” in \textit{IEEE International Symposium
			on Information Theory (ISIT),2019}
		
	
		\bibitem{SZG}
		C. Shangguan, Y. Zhang and G. Ge, ``Centralized Coded Caching Schemes: A Hypergraph Theoretical Approach," in \textit{ IEEE Transactions on Information Theory}, vol. 64, no. 8, pp. 5755-5766, Aug. 2018
		\bibitem{Stinson}
		D. R. Stinson, \textit{Combinatorial Designs$:$ Constructions and Analysis}. Springer, 2003.
		\bibitem{KrP1} S. Agrawal, K. V. Sushena Sree and P. Krishnan, "Coded Caching based on Combinatorial Designs," 2019 IEEE International Symposium on Information Theory (ISIT), Paris, France, 2019, pp. 1227-1231.
	
	    \bibitem{CJYT}
	    Minquan Cheng , Jing Jiang , Qifa Yan , and Xiaohu Tang
	    ``Constructions of Coded Caching Schemes
	    With Flexible Memory Size," in \textit{IEEE Transactions on Communications}, vol. 67, no. 6, pp. 4166-4176, June 2019
	\end{thebibliography}
\end{document}